\documentclass{svproc}
\usepackage{subfigure}
\usepackage{slashed}
\usepackage{amsmath,amssymb}
\usepackage{graphicx}
\usepackage{units}
\usepackage{bbold}
\usepackage{xcolor}
\usepackage{dsfont}
\usepackage[hyperfootnotes=false,colorlinks,citecolor=blue]{hyperref}
\usepackage{comment}
\usepackage[normalem]{ulem} 
\usepackage{comment}
\usepackage{autobreak}
\usepackage{wrapfig}
\usepackage{cite}
\usepackage{multicol}

\begin{document}
\mainmatter             
\title{Probing New Physics with \\
Multi-Messenger Astronomy}
\titlerunning{New Physics with Multi-Messenger Astronomy}  
\author{P. S. Bhupal Dev}
\authorrunning{P.S.B. Dev} 
\institute{Department of Physics and McDonnell Center for the Space Sciences, \\Washington University, St.~Louis, MO 63130, USA\\
\email{bdev@wustl.edu}
}

\maketitle              % 

\begin{abstract}
The burgeoning field of multi-messenger astronomy is poised to revolutionize  our understanding of the most enigmatic astrophysical phenomena in the Universe. At the same time, it has opened a new window of opportunity to probe various particle physics phenomena. This is illustrated here with a few example new physics scenarios, namely, decaying heavy dark matter, pseudo-Dirac neutrinos and light dark sector physics, for which new constraints are derived using recent multi-messenger observations. 
\keywords{Multi-messenger Astronomy, Neutrinos, Dark Sector}
\end{abstract}
\section{Introduction}
\begin{wrapfigure}{r}{0.45\textwidth}
\centering
  \vspace{-1.5cm}    \includegraphics[width=0.44\textwidth]{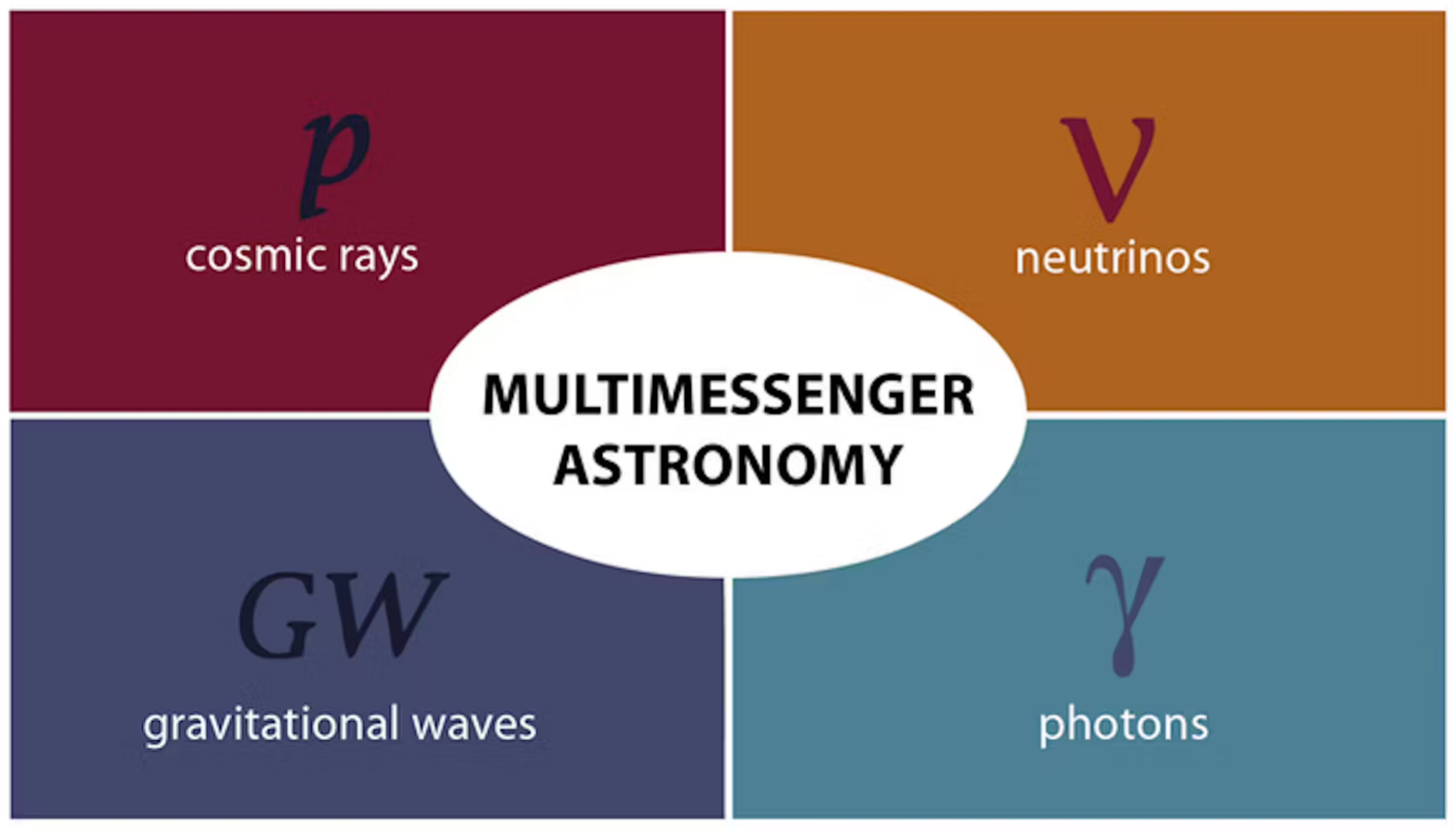}
    \vspace{-1cm}
    \caption{The four pillars of multi-messenger astronomy~\cite{10.1088/2514-3433/ac087ech1}.}
    \label{fig:multimes}
    \vspace{-0.5cm}
\end{wrapfigure}
Multi-messenger astronomy~\cite{10.1088/2514-3433/ac087ech1,Meszaros:2019xej}, a long-anticipated extension to the traditional optical and multi-wavelength astronomy, has emerged over the past decade as a transformative approach providing valuable new insights into energetic astrophysical phenomena. These insights arise from leveraging the inherently complementary information carried by different cosmic messengers (see Fig.~\ref{fig:multimes}), namely,  photons, cosmic rays (CRs), neutrinos and gravitational waves (GWs), about individual cosmic sources and source populations. In particular, two recent breakthrough discoveries, namely, the identification of high-energy neutrino (HEN) point sources~\cite{IceCube:2018cha,IceCube:2022der} and the multi-messenger observations of a binary neutron star (NS) merger event GW170817~\cite{LIGOScientific:2017vwq,LIGOScientific:2017ync}, have ushered in a new era of multi-messenger astronomy. 
The purpose of this proceeding, based on a plenary talk at PPC2024~\cite{dev_talk}, is to show how these recent developments provide a great opportunity not only for astrophysics, but also for particle physics. %Specifically, we will illustrate with a few examples how the multi-messenger observations of neutrinos, $\gamma$-rays and gravitational waves (GWs) can be used as new probes of beyond the Standard Model (BSM) physics. 
\section{HENs and Multi-Messenger Connection}
\begin{figure}[t!]
\centering
    \includegraphics[width=0.99\textwidth]{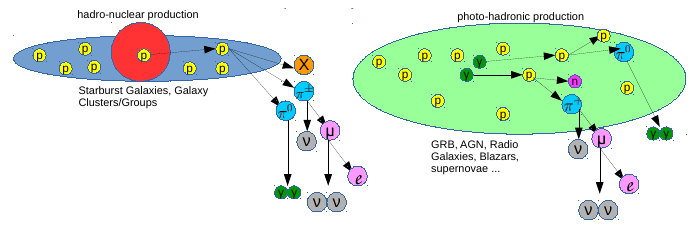}
    \caption{HEN production in astrophysical environments. %Accelerated CR protons  interact with ambient photons ($p\gamma$) or cold gas ($pp$), leading to the production of charged and neutral pions, which decay promptly to produce neutrinos and $\gamma$-rays.
    }
    \vspace{-0.5cm}
    \label{fig:HEN}
\end{figure}

The observation of HENs in the TeV-PeV range at the IceCube Neutrino Observatory~\cite{IceCube:2013cdw,IceCube:2013low} has commenced a new era in Neutrino Astrophysics~\cite{Arguelles:2024xkx}. It reveals a profound multi-messenger connection between high-energy CRs, neutrinos and $\gamma$-rays. In astrophysical environments with relativistic particle flows, some of the gravitational energy released in the accretion of matter can help accelerate protons and heavier nuclei to ultra-relativistic energies through diffusive shock acceleration or magnetic reconnection processes~\cite{Bell:2013vxa}. These accelerated CRs interact with ambient matter ($pp$) or radiation ($p\gamma$) to produce charged/neutral pions (and kaons at higher energies~\cite{Asano:2006zzb}). Charged pions decay into neutrinos: $\pi^+\to \mu^+\nu_\mu$, followed by $\mu^+\to e^+\bar{\nu}_\mu\nu_e$, while $\pi^0\to \gamma\gamma$ (see Fig.~\ref{fig:HEN}). Since a $\pi^0$ decays into two photons for every charged pion producing a $\nu_\mu\bar{\nu}_\mu$ pair, it gives a powerful relation between the $\gamma$-ray and neutrino fluxes~\cite{Ahlers:2015lln, Meszaros:2017fcs}: 
$E^2_\gamma\frac{dN_\gamma}{dE_\gamma}\approx \frac{4}{K_\pi}\frac{1}{3}E^2_\nu\left.\frac{dN_\nu}{dE_\nu}\right|_{E_\nu=E_\gamma/2}$, where $K_\pi\approx 2 (1)$ is the ratio of charged to neutral pions produced in $pp$ ($p\gamma$) interactions. Also, from the fact that the primary CR proton's mean relative energy loss per interaction, called the inelasticity, is $\kappa_{pp(p\gamma)}\approx 0.5 (0.2)$, the mean energy of the resulting neutrinos and $\gamma$-rays is $\sim 0.05$ and $\sim 0.1$ of the initial CR proton energy~\cite{Murase:2016gly}. 

Once produced, the weakly interacting neutrinos escape the astrophysical environment mostly unhindered, and just undergo vacuum oscillations before reaching the Earth with a roughly equal flavor ratio of $(1:1:1)$~\cite{Learned:1994wg}. The flavor ratio could be altered if e.g., the muon from pion decay rapidly loses energy due to environmental interactions such as synchrotron emission before decaying
(muon-damped case)~\cite{Rachen:1998fd}. Also, if the source is extremely dense with column density much larger than the inverse Thomson cross-section $\sigma_T^{-1}\sim 1.5\times 10^{24}~{\rm cm}^{-2}$, it could induce strong matter effect on the neutrinos~\cite{Dev:2023znd}. 

While efficient neutrino production thrives in a dense target field, the same dense field can be responsible for
the absorption of $\gamma$-rays inside the source, leading to the production of secondaries, which cascade down in energy until the source becomes transparent to them. This redistributed photon energy from the TeV–PeV range appears in the X-ray to multi-GeV $\gamma$-ray band, 
where observational efforts are focused to constrain the expected TeV-PeV HEN flux from a given 
source~\cite{Meszaros:2019xej, Arguelles:2024xkx}. 
In addition, the extragalactic background light further depletes the $\gamma$-ray flux at energies of ${\cal O}$(100 GeV)
and above depending on the redshift~\cite{Finke:2009xi}, thus further complicating the
simultaneous detection of TeV $\gamma$-rays produced alongside the TeV neutrinos from the same source. 

Nevertheless, on the particle physics front, the intimate connection between the HENs and $\gamma$-rays can be used to constrain various beyond-the-Standard Model (BSM) scenarios. Here we give two such examples to illustrate our point.  

\subsection{Decaying Heavy Dark Matter}
There is overwhelming evidence from astrophysical and cosmological observations for the existence of Dark Matter (DM), which constitutes 27\% of the energy budget (or 85\% total mass) of our Universe~\cite{Planck:2018vyg}. But the nature and properties of DM are still unknown, which are among the most important open questions in physics. If the DM thermalizes with the SM particles in the early Universe, there is a well-known partial-wave unitarity upper bound of {\cal O}(100) TeV on its mass~\cite{Griest:1989wd, Smirnov:2019ngs}. However, there exist ways to push the unitarity limit to much higher DM masses~\cite{Smirnov:2019ngs,Bernal:2023ura}.  Moreover, the DM need not be thermally produced nor absolutely stable. This allows us to entertain the possibility of a heavy decaying DM with mass in excess of ${\cal O}$(100 TeV), for which some of the strongest constraints actually come from the multi-messenger observations of HENs and very-high-energy $\gamma$-rays~\cite{Sui:2018bbh,Arguelles:2022nbl,Das:2023wtk,Fiorillo:2023clw,Song:2023xdk}. Interestingly, there is a slight tension between the IceCube measurements of the astrophysical flux using muon neutrinos and the starting event measurements, which can be explained by a decaying DM component, while being consistent with current neutrino and $\gamma$-ray flux limits~\cite{Chianese:2016opp,Sui:2018bbh, Skrzypek:2022hpy}. 

\subsection{Pseudo-Dirac Neutrinos}
The observation of neutrino oscillations indicates that neutrinos are massive; however, the nature of neutrino mass (Dirac or Majorana) remains an open question. Theoretically, it is also possible that neutrinos are pseudo-Dirac~\cite{Wolfenstein:1981kw,Petcov:1982ya,Valle:1983dk,Doi:1983wu,Kobayashi:2000md}, which are fundamentally Majorana, but essentially act like Dirac particles in most experimental settings due to extremely small active-sterile mass squared splitting $\delta m^2$. The theoretical and model-building aspects of pseudo-Dirac neutrinos have been extensively discussed in the literature. 
%; see e.g., Refs.~\cite{Chang:1999pb, Nir:2000xn,Joshipura:2000ts,Lindner:2001hr, Balaji:2001fi,Stephenson:2004wv,McDonald:2004qx, deGouvea:2009fp, Ahn:2016hhq, Joshipura:2013yba, Babu:2022ikf, Carloni:2022cqz}. 
In fact, in any model where neutrinos start as
Dirac particles with naturally small masses 
would receive quantum gravity corrections via higher-dimensional operators suppressed by the Planck scale making them pseudo-Dirac particles at a more fundamental level.
It is interesting to note that certain string landscape (swampland) constructions also prefer (pseudo)Dirac neutrinos over Majorana ones~\cite{Ooguri:2016pdq, Casas:2024clw}. 
Small $\delta m^2$ values could also be linked to the observed baryon asymmetry of the Universe~\cite{Ahn:2016hhq, Fong:2020smz}. Pseudo-Dirac neutrinos could also resolve the excess radio background anomaly~\cite{Chianese:2018luo, Dev:2023wel}. 

Irrespective of these theoretical motivations, the only experimental way to probe the active-sterile oscillations of pseudo-Dirac neutrinos is by going to extremely long baselines, which is possible with astrophysical sources of neutrinos~\cite{Beacom:2003eu, Keranen:2003xd}. 
In fact, until recently, solar neutrino data provided the most stringent upper limit on $\delta m^2_{1,2}\lesssim 10^{-12}~{\rm eV}^2$~\cite{deGouvea:2009fp, Ansarifard:2022kvy},  better than the Big Bang Nucleosynthesis limit on $\delta m^2_i\lesssim 10^{-8}~{\rm eV}^2$~\cite{Barbieri:1989ti, Enqvist:1990ek} for maximal mixing.

\begin{wrapfigure}{r}{0.45\textwidth}
\centering
     \includegraphics[width=0.45\textwidth]{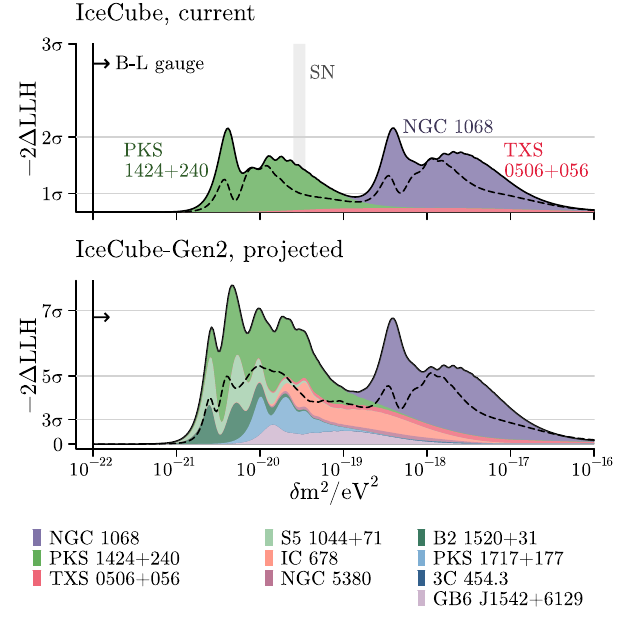}
      \vspace{-0.5cm}
    \caption{IceCube constraint (top) and Gen2 sensitivity (bottom) on the active-sterile neutrino mass splitting for the pseudo-Dirac scenario~\cite{Carloni:2022cqz}.}
    \label{fig:pseudoDirac}
    \vspace{-0.5cm}
\end{wrapfigure}
 
The recent multi-messenger identification of a handful of point sources for astrophysical neutrinos~\cite{IceCube:2022der} allowed us to set the first IceCube limits on the pseudo-Dirac neutrino hypothesis in the $\delta m^2_i\in [10^{-21},10^{-16}]~{\rm eV}^2$ range~\cite{Carloni:2022cqz} (see also~\cite{Rink:2022nvw, Dixit:2024ldv}), as shown in Fig.~\ref{fig:pseudoDirac}. The basic idea is that the oscillation probability is sensitive to the combination $\delta m^2 L/E_\nu$; so once the distance $L$ to the source is known from multi-messenger observations, and the neutrino flux from this source is measured as a function of energy $E_\nu$, we can probe certain $\delta m^2$ values. If the active neutrino oscillates into the sterile component as it reaches the Earth, it will give rise to a flux deficit, which will be energy-dependent, and hence, robust against astrophysical flux uncertainties.  

Ref.~\cite{Carloni:2022cqz} only used the IceCube track-like sample (mostly involving muon neutrinos, with a small fraction coming from tau-induced tracks), and hence, was insensitive to the full neutrino flavor information. It was subsequently shown~\cite{Fong:2024mqz, Dev:2024yrg} that the standard flavor triangle predictions could also get modified in presence of pseudo-Dirac neutrinos, which might be observable in future neutrino telescopes like IceCube-Gen2 and KM3NeT.

\section{Multi-Messenger Studies of GW170817}

\begin{wrapfigure}{r}{0.45\textwidth}
\centering
\vspace{-0.8cm}
\includegraphics[width=0.45\textwidth]{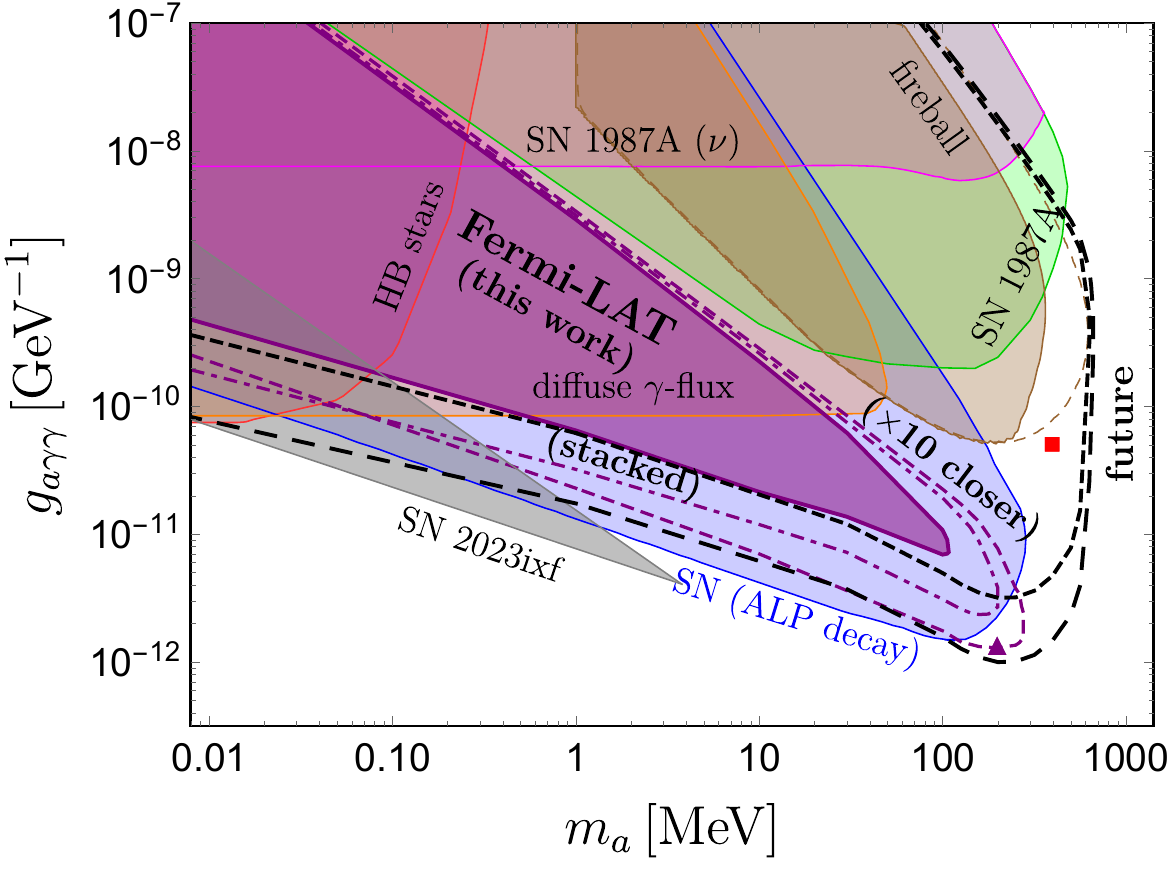}
\vspace{-0.8cm}
\caption{Current exclusion (shaded) and future sensitivity contours in the $(m_a,g_{a\gamma\gamma})$ plane~\cite{Dev:2023hax}. 
}
\label{fig:merger}
\vspace{-0.5cm}
\end{wrapfigure}
The extreme astrophysical environments in the vicinity of compact objects like black holes, NSs, magnetars,
binary black hole and NS mergers have recently emerged as 
a new tool for probing light dark-sector physics, complementary to and beyond the traditional arena of stellar and supernova environments~\cite{Raffelt:1996wa}. Much of this recent progress is driven by  data from across the electromagnetic spectrum, as well as neutrinos and  GWs, together with their multi-messenger studies. In particular, the multi-messenger discovery of the NS merger event GW170817~\cite{LIGOScientific:2017vwq,LIGOScientific:2017ync} has opened a new window to BSM particle searches, such as axions and axion-like particles (ALPs)~\cite{Dietrich:2019shr,Harris:2020qim, Zhang:2021mks,Fiorillo:2022piv,  Dev:2023hax,Diamond:2023cto}, CP-even scalars~\cite{Dev:2021kje}, and dark photons~\cite{Diamond:2021ekg}. Fig.~\ref{fig:merger}  illustrates the result for ALPs. The basic idea is to produce the ALP (or any dark sector particle) in the NS merger environment, which subsequently decays into observable photons. By analyzing the spectral and temporal features of the photon signal induced by ALP decay, we thus derive the first multi-messenger constraints from GW170817 on the ALP-photon coupling. Future MeV $\gamma$-ray missions such as e-ASTROGAM and AMEGO-X could significantly improve the sensitivity, going beyond the current exclusion from SN1987A multi-messenger data~\cite{Muller:2023vjm}. In particular, electromagnetic observations of the NS merger within the first second of the GW detection (possible with the early-warning system~\cite{Sachdev:2020lfd}) would be crucial to isolate the ALP-induced signal from any potential astrophysical background.  

\section{Conclusion and Outlook}
New messengers lead to new insights. This is an exciting era of multi-messenger astronomy -- the coordinated observation and interpretation of multiple signals received from the same astronomical event. While still in its infancy, it holds great promise for both astrophysics and particle physics communities. The multi-messenger perspective is already yielding more than just the sum of its parts -- and we can expect to see more profound discoveries in the future. Here we illustrated how the recent multi-messenger observations with photons, neutrinos and GWs have opened new windows of opportunity into the BSM world. 
 
%
%\section*{Acknowledgments} 
\medskip
\noindent
{\bf Acknowledgments:} The work of B.D. was partly supported by the U.S. Department of Energy under grant No. DE-SC0017987.
%

%\begin{multicols}{2}
\bibliographystyle{utphys}
\bibliography{ref_dev.bib}
%\end{multicols}

%
\end{document}